# A Framework for Computing on Large Dynamic Graphs

Zhao Yu Dong

## Abstract

This proposal presents a graph computing framework intending to support both online and offline computing on large dynamic graphs efficiently. The framework proposes a new data model to support rich evolving vertex and edge data types. It employs a replica-coherence protocol to improve data locality thus can adapt to data access patterns of different algorithms. A new computing model called protocol dataflow is proposed to implement and integrate various programming models for both online and offline computing on large dynamic graphs.

A central topic of the proposal is also the analysis of large real dynamic graphs using our proposed framework. Our goal is to calculate the temporal patterns and properties which emerge when the large graphs keep evolving. Thus we can evaluate the capability of the proposed framework.

# 1. Introduction

Dynamic big graphs are arising from a lot of applications, ranging from online social business and communication networks to financial transaction networks, citation network, gene regulatory networks and disease transmission networks. Those graphs keep evolving and this makes computing on them more difficult than static graphs. A computing framework supporting dynamic graphs can be theoretically and practically important. For example in a social network it can help us to answer questions like "who makes the most friends this month?" or "who is the most active person in my friends?". Some theoretical research has been done on dynamically evolving graphs to model the graph phenomena ([Jurij2008], [Andre2013], [Manuel2013]). Those studies lead to the methods to learn and mine the dynamic graphs more efficiently and help to capture the emerging patterns when the graphs evolve, such as community detection [Nicola2013], link prediction [Ole2013], cascade prediction [Wojciech2010]. But if we apply those algorithm to a real word dynamic graph and perform on time computing, challenges will arise. We summarize them as following.

- **Heterogeneous evolving data types**

Many applications need to handle heterogeneous data type. For example in a citation graph, there will be authors and papers. Each paper can have multiple authors and it can refer to other papers. The heterogeneity of data makes data schema plays an important role when mining such graphs. Those schemas give the data scientist a hint of some insight of the data and can help them to design good learning algorithms. For example in [Yizhou2012] the author proposes algorithms to mine the heterogeneous graph data using data schema. Trinity [Bin2013] is a graph computing system allowing the user to implicitly define data of different types and expose the data schemas to the application. For dynamic graphs, the data type can evolve over time. User may need to introduce new type of node and links, current schema support is not suitable for data type evolving.

- **Ad hoc data access pattern**

Graph nodes and links are highly irregular and data access pattern in graph exploration is very ad hoc. For dynamic graphs this problem will be particularly acute since the access pattern and graph structure itself may change very rapidly. The computing performed on graphs can be broadly divided into two categories, one is online computing and the other is offline graph analytics. Online computing requires low latency and quick response, such as reachability queries [Ruoming2008] and key word search over graphs [Konstantin2008]. Offline graph analytics typically requires high throughput, such as subgraph pattern matching ([Zhao2012] [Danai2011]) and data mining on graphs [UKang2009]. Those two kinds of computing often have different data access patterns. Online query tends to rely on super-linear indices and some other sketches which are often computed by analytical jobs [Bin2012]. Improving locality is considered a promising way to improve the performance when ad hoc data access patterns present in large scale graph computing. In [Josep2010] a heuristic algorithm is proposed to enforce the local semantic (every graph node along with all its directed neighbors will be co-located in the same partition). In [Jayanta2012] the authors use fairness requirement to guide the replication decision and leverage predictive models of the data access pattern to manage the data replicas. [Eiko2013] uses pre-fetch strategy and edge-centric view to improve the locality. Although a lot research has been done on this, ad hoc data access pattern still is a hinter to scale the graph computing to large dynamic data sets. A single method can not solve the problems we have when we conduct a variety of computing on large dynamic graphs. With the graph changing dramatically, load balance, network communication, replica overhead and stability should be all considered when managing dynamic data.

- **Computing with constantly incoming mutations**

The ability to process online mutation on a large volume of data is very important for dynamic graph computing framework. In a real dynamic graph, a stream of mutation will be applied to the graph and can make the graph evolve consistently. Currently Spark Streaming [Mateiz2012] and Naiad ([Frank2012] [Frank2013]) support the user to apply a sequence of mutations to the large graph. But this still is not enough to fulfill the demand of computing on dynamic graphs. For example, graph researchers may want to know how a pattern in dynamic graph emerges or how a community comes to be. They need to design algorithms to mining the graph evolvement process. So the systems should support the computing both on graph evolving trace and graph snapshot on a specified time point. Computing like this needs to adapt to the graph changes first and then reschedule the computing on the entire graph.

So far, many computing frameworks have been proposed to address the problems encountered when perform computing on large graphs, such as Pregel [Grzegorz2010], PowerGraph [Joseph2012], Graphlab [Yucheng2012], Trinity [Bin2013], Naiad [Frank2012], GraphX [Reynold2013], Kineograph [Reymond2012] and GPS [Semih2013]. Those all partially solve the problems we facing on dynamic graph computing. In this proposal we will design and built a computing framework to fully support computing on dynamic graphs. We attempt to make following contributions:

1) Introduce a new data model to support evolving heterogeneous data types for large dynamic graph computing (*section 2.1*).
2) Implement a replica-coherence protocol based data manager to improve communication performance, load balance, and alleviate the impact of poor locality on large dynamic graphs (*section 2.2*).
3) Design and build a computing model which can efficiently integrate online and offline computing on large dynamic graphs by supporting data sharing, asynchronous computing, various programming models and application-specific scheduling (*section 2.3*).

The remaining of the proposal will be constructed as following: Section 2 presents our proposed solution, provide the relevant aspects and define the work scope. Section 3 outlines how we intend to provide an empirical validation of our proposed solution.

# 2. Proposed solution

The complexity of data and the supported computing mainly determine the complexity of distributed system. The proposed solution here will focus on the data complexity and how to support both online and offline computing efficiently.

## 2.1 Data model

Graphs from different area will have different schemas. And the node and link schema can still evolve. To support evolving schema in this proposal, the nodes and links in a graph are seen as abstract entities. An application can attach any schema to the nodes and links according their real needs. If a graph has no any schema attached to it, it is called an abstract graph. If some schema is attached to nodes or links of a graph, it is called schematized graph. The proposed data model will expose features to applications to declare schemas with sufficient support for evolving. To keep track of the evolving schema, a version number is introduced to schema declaration. A schema can have different versions. This makes the schema declaration looks like a template in object oriented language.

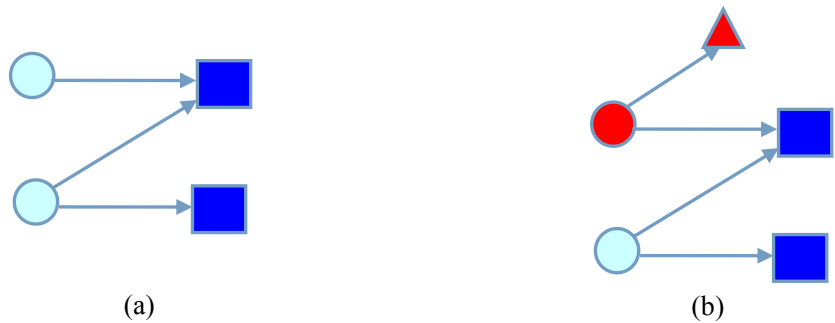

(a)          (b)

Figure 1. A circle stands for an author. A square stands for a paper. A triangular stands for a school. (a) is the graph with author and paper nodes. (b) shows the graph evolves and a new type of node (school node) is added. And the author is changed to add a new link to the school he belongs to.

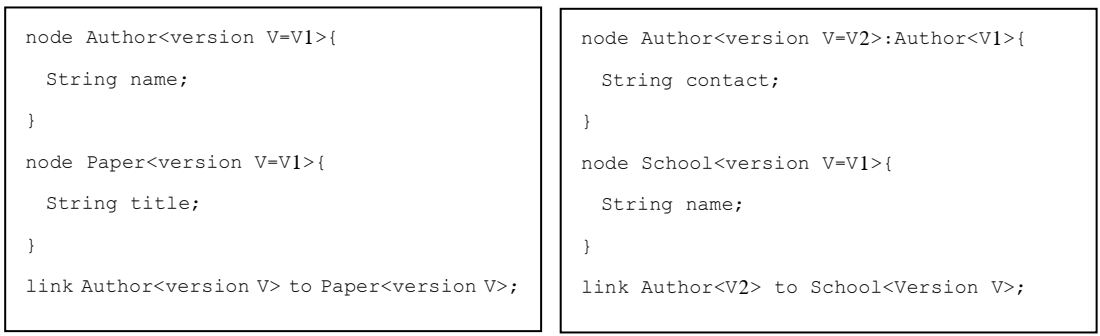

```
node Author<version V=V1>{
  String name;
}
node Paper<version V=V1>{
  String title;
}
link Author<version V> to Paper<version V>;
```

```
node Author<version V=V2>:Author<V1>{
  String contact;
}
node School<version V=V1>{
  String name;
}
link Author<V2> to School<Version V>;
```

Figure 2 Left side is the schema declaration of graph (a) in figure 1. Right side is the evolved schema. Node author has a new version (version V2) by inheritance from version V1. A new node type school is introduced and author of version V2 is connected to it.

Figure 1 and 2 are examples showing how a author-paper graph may evolve and its associated schema declarations. Exposing the version number of schema can enable the applications to conduct different computation according different versions. Or the same computation can be conducted on a set of versions. This is needed for mining the temporal patterns.

## 2.2 Data management

Large graphs are partitioned and stored across a cluster which may contain thousands of machines. Static partition of graph is not suitable for dynamic graphs. Since the graph can change rapidly and load balance can be broken thus make the performance deteriorate. How to dynamically partition the evolving graph will have a big impact on the overall computing performance. The central problems here is when a change is applied on the graph, how to adjust the partition. A graph partition is mainly determined by two factor: load balance and communication latency. This proposal will developer sophisticate models to measure the effect a change in graph will have on those two factors. A scheduler will be implemented with this model to achieve dynamic equilibrium between those two factors.

To optimize the performance, data replicas are needed to achieve locality semantics. In contrast to static graphs, replicas of dynamic graph need to maintain consistency. When a change is applied on one replica, this changes should also be spreaded to other replicas. The consistency can be achieved with Paxos state machine [Leslie1998]. Another problem is the distribution of those replicas need to be scheduled. This is to mitigate the communication overhead. If data of a replica on one machine is heavily requested by another machine. This replica should be swapped to the requester's machine. To improve locality semantics and consistency, in this proposal a distributed replica-coherence protocol will be designed to manage the replicas efficiently. With this protocol replicas can be created and redistributed over the cluster while maintaining the consistency. Another problem is how to collect the obsolete replicas. In this proposal various strategies will be explored to deal with this problem.

## 2.3 Computing on dynamic graph

This proposal is mainly to support online and offline computing on dynamic graphs and integrate those two types of computing efficiently. We aim to achieve:

1) **Data from online computing and offline computing can be shared.**
2) **Online and offline computing with various programming models can be integrated.**
3) **Online and offline computing can be conducted asynchronously.**
4) **Fine-grained application-specific scheduling of online and offline computing.**

This section will illustrate the proposed computation framework. First it describes the versioned data set. Next it introduces the distributed view which is used to enable the data sharing. Finally based on this it presents the protocol dataflow model.

## 2.3.1 Versioned data set and snapshot

In our proposed system all data items will have its own version number. Only when a new mutation is applied to a specific data item, can a new version be created. The data version consists of two parts: the first part is the epoch identifier and the second part is version number within its owning epoch (as shown in figure 3(a)).

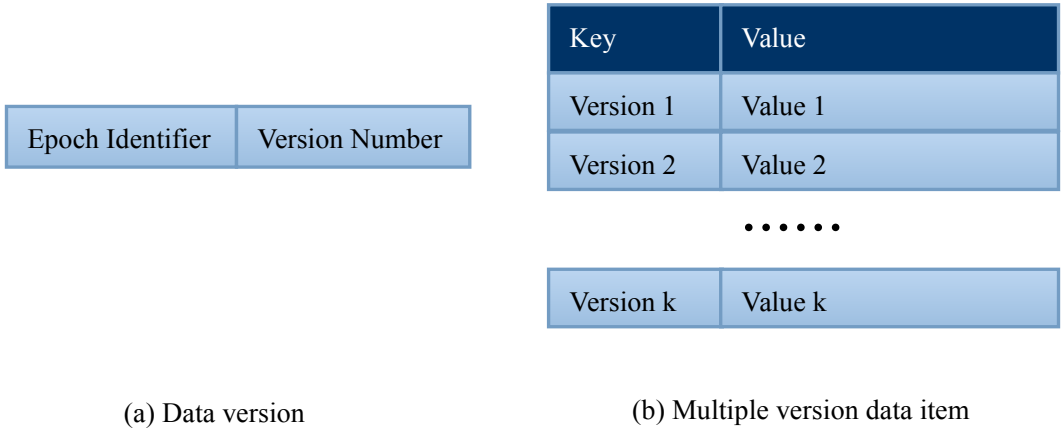

(a) Data version          (b) Multiple version data item

Figure 3 Versioned dataset

Most offline computing is assumed to be conducted on a specific snapshot. Since each mutation in a specific epoch has its own version of data created. A stream of mutation will create a data set with multiple data version, each version of data item will be stored as a key-value pair (Figure 3(b)). A snapshot can be described as a data set with the data version it includes.

$$snapshot(v) = \{d(i_v)\},$$
$$i_v = \max_{v' \leq v}\{v'\}$$

Here $d(i_v)$ is the value of data item $d$ with version $i_v$. The version $i_v$ equals the max version of all version numbers less than $v$.

The incoming mutations are dispatched from the ingest node to the data node asynchronously. So the snapshots will also be created asynchronously in the data node. To track the progress of global snapshots, the proposed framework will use a Paxos [Leslie1998] based protocol. Figure 4 shows the configure for global snapshot progress tracking. All the data nodes have their local snapshots and their progress is ahead of the global snapshot. Some systems like Kineograph [Reymond2012] uses a central snapshooter to track the global snapshot. All the mutations in an epoch will be delayed if the global snapshot from the previous epoches are not finished. In our system mutations from different epochs can be dispatched concurrently. When dispatching a mutation the ingest node first will look up the data node which contains the data the mutation will apply. If the local snapshot of previous epoches have been defined in those data nodes, the mutation can be safely dispatched. There is no need to wait until the global snapshot is defined.

The computing and incoming mutations are synchronized to maintain consistency. When a computing is scheduled on the graph, it will specify the snapshots which the computing will be conducted on. The computing is launched until all the global snapshot it will process become available. This is important because some application need this to guarantee correctness.

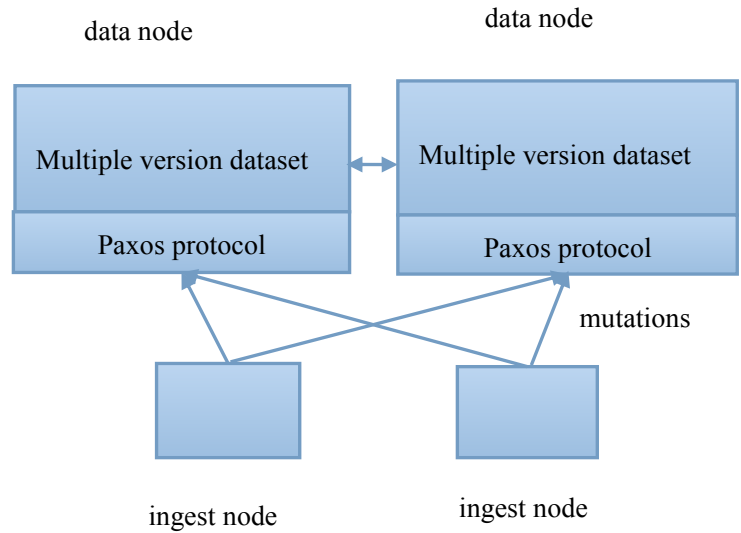

Figure 4 global snapshot progress tracking

## 2.3.2 Distributed view

To efficiently share data between online computing and offline computing, a data abstraction is needed to address the data set and computations can be expressed in terms of the data abstraction. Here we borrow the view concept from DBMS. In traditional database systems a view is an abstraction of the data set returned from a SQL query. The view can enable the data sharing between different SQL queries. In the proposal we will use distributed view to address the data from a computation. A distributed view is immutable and is expressed by the computation from which it is generated. This is like the lineage in a RDD [Matei2012]. So the fault-tolerance of a distributed view can be achieved by tracing back its lineage and redo the computations along the lineage path. The distributed view abstraction is mainly to support the asynchronous execution on the data set.

## 2.3.3 Protocol Dataflow

Protocol dataflow is the framework proposed to efficiently integrate online and offline graph computing. This section will describe the structure of protocol dataflow, and then present the optimization for graph computing.

### 2.3.3.1 Structure of protocol dataflow

Protocol dataflow is constructed as a directed graph. Computing on it starts with an ingress vertex and ends with an egress vertex. The ingress vertex receives input from external source and encapsulates the data into messages according to a predefined protocol. The egress vertex receives a sequence of message, decapsulate them and send the data to an external consumer. The ingress and egress vertex separate the context of computing on protocol dataflow and make it easy to embedded protocol dataflow in other computing environments.

In protocol dataflow vertices enclosed in the ingress and egress vertex are called internal vertex. An internal vertex can have several input queues and output queues. Each input queue can receive a sequence of messages from another vertex. The computation in the vertex will parse the messages, do computation with the data in the messages and emit the result messages to output queues. A vertex can have two schedulers, one scheduler works on the input queues and another works on the output queues. The input scheduler allows the computing on this vertex to employ the application-specific scheduling to improve performance. The output scheduler provides a chance to optimize the communication between the neighbouring vetices. The structure of vertex in protocol dataflow is shown in Figure 5.

Protocol dataflow is general enough to be used to implement different programming models by defining different protocols. A protocol is defined by two parts. First is the format of the messages each vertex processes. Second is the semantics that defines the computation each vertex dose. Programming models defined by different protocols can be integrated together as shown in Figure 6.

Computing done on protocol dataflow resembles what is done on a communication network. A network has a graph like topology and uses protocols to manage how each node to communicate with others asynchronously. Vertices in protocol dataflow are analogous to nodes in networks. The vertices are stateful and are more complex than the vertices in MapRuduce-like dataflows. Data-centric computations on dataflows need to deal with objects from two spaces: data space and coordination space [Derek2011]. In protocol dataflow the information of data space and coordination space and encoded together in messages. The semantics of the protocol defines how each vertex does computation upon the received messages. So the control flow in protocol dataflow is data-dependent and there is no need to use central scheduler to guarantee the correctness of computations. This will be easier for protocol dataflow to scale out to large data centers [Kay2013].

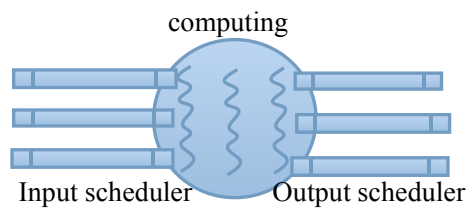

Figure 5 A vertex in protocol dataflow can have several input queues and output queues.
An input scheduler works on the input queues and a output scheduler works on the
output queues.

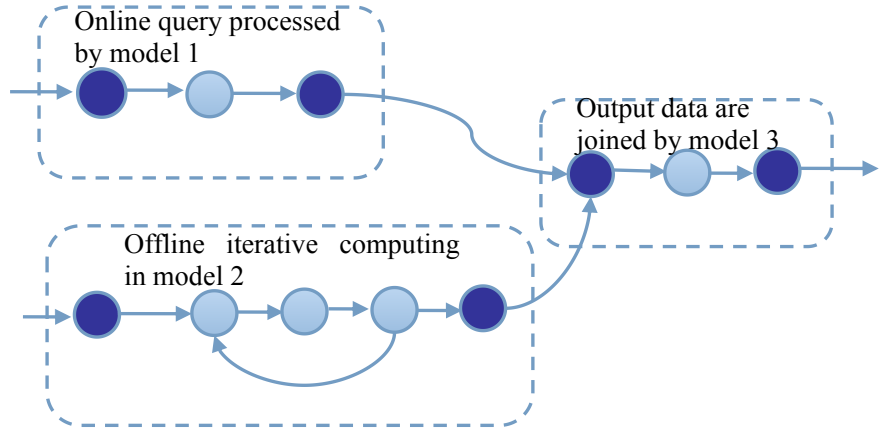

Figure 6 Different programming models of protocol dataflow can be integrated.
The dark circle are ingress vertices and egress vertices. Dashed rectangle
stands for the computing in different models.

## 2.3.3.2 Event delivery

To observe the execution of programs on protocol dataflow, a mechanism is needed to deliver some events to the external observers. The events should be delivered in an order which is meaningful and maintain integrity of a predefined constrains. Those events will be important for the observer to reason the execution and synchronize other actions. For example a word count program may send an event when a line in the input text is finished. The event observer receives the event and will print a message to the output queue. Protocol dataflow allows the user to define any kind of event and it provides interface to declare observers for the events they interest in. A key problem in delivering the event is to keep the causal-effect relation in the events order. This causal-effect relation forms a partial order and imposes a constrain on the possible event sequence. Here we use "→" to denote the relation. For two events $e_1$ and $e_2$, if $e_1$ has a causal-effect to $e_2$ we denote it as $e_1 \rightarrow e_2$. If $e_1$ and $e_2$ are concurrent we denote it as $e_1 \leftrightarrow e_2$. To deliver the events correctly each vertice in protocol dataflow is assigned a logical clock. The time of the clock when event e occurs is denoted as $T(e)$. The time of the events should keep this condition:

$$\text{If } e_1 \rightarrow e_2 \text{, then } T(e_1) < T(e_2)$$

Here the relation $e_1 \rightarrow e_2$ is application specific. Different program models will have different criteria to determine whether an event can precede another event. So each program model implemented with protocol dataflow needs to register its own function to check the causal-effect relation with two given events. The algorithm to deliver the events while preserving the predefined "→" relation is a well-known study in [Leslie1978]. The benefit of this algorithm is it doesn't share global state and there is no need to employ a central event scheduler.

## 2.3.3.2 Graph computing on protocol dataflow

In graph computing each vertex needs to communicate with its neighboring vertices to update its own state. The communication between the neighboring vertices can be implemented with a join-group-by operator. To efficiently join the neighboring vertices for each snapshot of the graph data, we will use a join view to speed up the join-group-by operation. In the proposed research we will explore the structure of the join view and how join-group-by can be implemented with it.

Another way to optimize the graph computing is to use fine-grained scheduling. Protocol dataflow allows the user to implement different scheduling policies to take advantage of the application-specific optimization to improve the performance. For example the Dijkstr's single source shortest path algorithm can be accelerated with a priority queue. Procotol dataflow allows the users to adopt this policy to speed up the convergence of the computation.

Communication optimization is also very important for the performance. PowerGraph [Joseph2012] uses the vertex cut instead of edge cut policy to optimize the communication. Trinity [Bin2013] buffers the hub remote vertices on local machine to reduce the total message passing. In protocol dataflow we will explore various message scheduling policies to improve the communication performance.

## 2.3.4 Discussion

In this proposal the incoming mutations can be incorporated timely and the computation result can be available both for online and offline computing. A streaming processing system can handle the incoming mutations with low latency, like Storm [StormUrl], and TimeStream [Qian2013]. But those systems don't combine the online and offline processing efficiently. None of our four goals (*section 2.3*) can be achieved in those system. Spark Streaming [Mateiz2012] uses the discretized streams abstraction to model the continuous streaming computation as a series of stateless, deterministic batch computation on small time intervals. It can achieve the goal 1) and 2). But Spark Streaming is based on the RDD [Matei2012] abstraction, so far lacks the ability to process data flow asynchronously. It can not achieve the goal 3) and 4). Galios [Donald2013] allows the computation to be scheduled with a coarse-grained application-specific scheduler. But fine-grained scheduler can not be achieved. And Galios is mainly for static graph computing. Problems on dynamic graph computing are not well addressed. Naiad [Frank2012] uses the timely dataflow computation model and it can achieve the goal 1) and 3). But goal 2) and 4) are not addressed. The protocol dataflow is very general. It can be used to implement timely dataflow and other programming models like graph parallel models (i.e. Vertex-centric model [Yucheng2012], edge-centric model [Roy2013] and graph-centric model [Yuan2014]) and data parallel models (i.e. MapReduce [Jeffrey2004]). Different kinds of program models are needed by different computations for better performance. All those models can be implemented on top of protocol dataflow and can be integrated together. So a specific application in protocol dataflow can be divided into a serial of computations and each computation can choose the program model which can achieve the most performance gain. Protocol dataflow servers as a common runtime for different program models.

## 2.4 Framework architecture

The proposed architecture is similar to Lambda architecture [Nathan2013]. The system is composed by multilayer to keep simplicity, meanwhile is fault tolerant, robust and scalable. Figure 7 shows the overall architecture of the proposed solution and we will build the system on top of a distributed memory cloud. The problems encountered in dynamic graph computing are solved in the corresponding layer with our proposed solution.

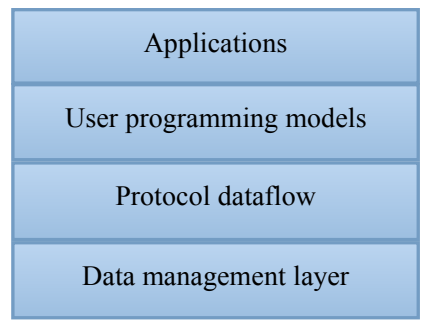

Figure 7    Architecture for dynamic graph computing framework

## 3. Evaluation Plan

The evaluation plan consists of four stages:
1. In the first stage, we will implement a prototype of the protocol dataflow on a single machine and verify its ability to support other programming models. Then we will extend this implementation to a distributed cluster and design the fault-tolerance mechanism.
2. In second stage we will implement the storage layer in the proposed framework. Then design and implement the join view and optimization to support efficient large dynamic graph computing.
3. In stage three we will implement some existing programming models with protocol dataflow and evaluate the system in terms of those aspects: 1) performance of online computing; 2) performance of offline computing; 3) ability to timely incorporate incoming mutations; 4) efficiency of data management layer. Based on the evaluation we will tune and optimize the system.
4. Finally we will use our system to mine some real dynamic graphs to assess how well it can capture the temporal patterns emerging when the graphs keep evolving.